\documentclass{sig-alternate-2013}
\usepackage{balance}  
\usepackage{graphics} 
\usepackage{times}    
\usepackage{url}      
\usepackage[usenames,dvipsnames]{xcolor}

\usepackage{soul}

\def\pprw{8.5in}
\def\pprh{11in}

\usepackage[pdftex]{hyperref}
\hypersetup{
pdftitle={STFU NOOB! Predicting Crowdsourced Decisions on Toxic Behavior in Online Games},
pdfauthor={Jeremy Blackburn and Haewoon Kwak},
pdfkeywords={League of Legends; online video games; toxic behavior; crowdsourcing; machine learning},
bookmarksnumbered,
pdfstartview={FitH},
colorlinks,
citecolor=black,
filecolor=black,
linkcolor=black,
urlcolor=black,
breaklinks=true,
}

\makeatletter
\def\@copyrightspace{\relax}
\makeatother

\begin{document}



\title{STFU NOOB! Predicting Crowdsourced Decisions \\on Toxic Behavior in Online Games \\ {\LARGE [Please cite the WWW'14 version of this paper]}}

%
%
%
%
%

\numberofauthors{2} 
%

\author{
%
%
\alignauthor
Jeremy Blackburn\\
        \affaddr{University of South Florida}\\
        \affaddr{Tampa, FL, USA}\\
        \email{jhblackb@mail.usf.edu}
\alignauthor
Haewoon Kwak\\
        \affaddr{Telefonica Research}\\
        \affaddr{Barcelona, Spain}\\
        \email{kwak@tid.es}
}



\maketitle
\begin{abstract}

One problem facing players of competitive games is negative, or \emph{toxic}, behavior.
League of Legends, the largest eSport game, uses a crowdsourcing platform called the \emph{Tribunal} to judge whether a reported toxic player should be punished or not.
The Tribunal is a two stage system requiring reports from those players that directly observe toxic behavior, and human experts that review aggregated reports.
While this system has successfully dealt with the vague nature of toxic behavior by majority rules based on many votes, it naturally requires tremendous cost, time, and human efforts.

In this paper, we propose a supervised learning approach for predicting crowdsourced decisions on toxic behavior with large-scale labeled data collections; over 10 million user reports involved in 1.46 million toxic players and corresponding crowdsourced decisions.
Our result shows good performance in detecting overwhelmingly majority cases and predicting crowdsourced decisions on them.  
We demonstrate good portability of our classifier across regions.  Finally, we estimate the practical implications of our approach, potential cost savings and victim protection.

\end{abstract}

\category{K.4.2}{Computers and Society}{Social Issues}[Abuse and crime involving computers]
\category{J.4}{Computer Applications}{Social and Behavioral Sciences}[Sociology, Psychology]

\keywords{League of Legends; online video games; toxic behavior; crowdsourcing; machine learning}


\setcounter{section}{-1}

\section{Executive Summary}

``STFU NOOB!''
If you've played an online video game, there's a good chance you've heard this before.
Toxic behavior is a part of life in the world of modern mulitiplayer games, but does it have to be?
Do gamers need to be subjected to beratement for every little mistake?
If another player doesn't get his way, do his teammates have to sit and watch them actively try to sabotage the rest of the game?
If someone consistently exhibits this type of behavior, does the entire community just have to suffer?

As gaming has grown from simple two player games like PONG to a multi-billion dollar, so too has toxic behavior become more severe.
Toxic behavior poses a large threat to the gaming industry.
Estimates put griefing as the cause of about 25\% of calls to customer support lines.
Not only is this a huge cost to game operators, but, it demonstrates the kind of damage that toxic players can cause.
The stress caused by harassment and other forms of toxic behavior can cause players to become fatigued to the point that they quit the game.
With the advent of the free-to-play business model, where the game is given away for free and the developers are supported via micro-transactions, a game's sustainability is directly related to the strength of its community.
A large, healthy community attracts new players and keeps existing players engaged and willing to spend money.
Left unchecked, toxic behavior threatens to tear a game's community apart.

There have been many attempts to deal with toxic behavior.
The earliest required direct human intervention, usually from a game master/administrator.
While accurate and decisive, this type of system doesn't scale when there are 10s of millions of active players.
More recently, crowdsourced systems, which make use of the input of many humans, have arisen.
For example, the Overwatch system uses expert input to detect cheaters in Counter-Strike: Global Offensive.

Another example is the League of Legends Tribunal, which allows people to vote on whether an accused player is toxic.
While Overwatch deals with a somewhat black-and-white decision, the Tribunal deals with a more nebulously defined set of behaviors.
Maybe the player that appears to be deliberately rushing in and dying just needs to ``git gud.''
Maybe the player lashing out in chat just had a bad day and would normally never act like that.
The Tribunal collects multiple instances of potentially toxic behavior, collected from reports by players that actually experienced the behavior, and presents them to a panel of reviewers recruited from the community who cast votes for guilt or innocence.
The Tribunal has been deemed a success with claims that it results in reduced recidivism.
But, it still requires significant human effort, and it is slow by nature: a number of reports must be received before cases are even presented to reviewers.

In this paper, we analyze reports from several million League of Legends matches and their corresponding crowdsourced decisions from the Tribunal.
We find that Tribunal reviewers base their decisions on in-game performance of both the accused and the other players in the game, chat logs, and the information provided by in-game reports.
We then use machine learning techniques to build a model for detecting toxic behavior that successfully discriminates between guilty and innocent behavior about 80\% of the time, and between overwhelming agreement on innocence about 88\% of the time.

Our findings have broad impact.
There is an obvious contribution to the gaming industry, but, we also provide a deeper understanding of toxic behavior.
We show that, at least in certain scenarios, computers can be used to help find bad behavior.
Our experimentally derived model hints at the possibility for automatic detection of, for example, cyber bullying.
Ultimately, we demonstrate that machine learning, with a little help from the crowd, can tell if you are being toxic or just need to git gud.

\section{Introduction}\label{sec:introduction}

Bad behavior on the Internet, or toxic disinhibition~\cite{suler2004online}, is a growing concern.
The anonymity afforded by, and ubiquity of, online interactions has coincided with an alarming amount of bad behavior. 
Computer-mediated-communication (CMC), without presence of face-to-face communication, naturally leads to hostility and aggressiveness~\cite{ho2008social}.
In multi-player gaming, one of the most popular online activities, bad behavior is already pervasive.  
Over the past two decades, multi-player games have evolved past simple games like PONG, growing so popular as to spawn an ``eSports'' industry with professional players competing for millions of dollars in prize money.
Competition has been considered a good game design element for enjoyment, but at the same time, it leads to intra- and inter-group conflicts and naturally leads to aggressive, bad behavior.  

Usually, bad behavior in multi-player games is called \emph{toxic} because numerous players can be exposed to such behavior via games' reliance on player interactions and the damage it does to the community.
The impact of toxic players is problematic to the gaming industry.  For instance, a quarter of customer support calls to online game companies are complaints about toxic players~\cite{davies2011gamers}.  
However, sometimes the boundary of toxic playing is unclear~\cite{Chesney09,Foo04,Lin05,Mulligan03} because the expected behavior, customs, rules, or ethics are different across games~\cite{Warner05}.  
Across individuals, the perception of this grief inducing behavior is unique.  
Subjective perception of toxic playing makes toxic players themselves sometimes fail to recognize their behavior as toxic~\cite{Lin05}.
This inherently vague nature of toxic behavior opens research challenges to define, detect, and prevent toxic behavior in scalable manner. 

A prevalent solution for dealing with toxicity is to allow reporting of badly behaving players, taking action once a certain threshold of reports is met.
Unfortunately, such systems have flaws, both in practice and, perhaps more importantly, their perceived effectiveness by the player base.
For example, Dota 2, a popular multi-player developed by Valve, has a report system for abusive communication that automatically ``mutes'' players for a given period of time.
While the Dota 2 developers claim this has resulted in a significant drop in toxic communication, players report a different experience: not only does abusive communication still occur, but the report system is ironically used to grief innocent players\footnote{http://tinyurl.com/stfunub1}\footnote{http://tinyurl.com/stfunub2}\footnote{http://tinyurl.com/stfunub3}.

Riot Games, which operates one of the most popular competitive games, called \textit{League of Legends} (LoL), introduced the \emph{Tribunal} in May 2011.  As its name indicates, it uses the wisdom of crowds to judge guilt and innocence of players accused of toxic behavior; accused toxic players are put on trial with human jurors instead of automatically being punished.
While this system has successfully dealt with the vague nature of toxic behavior by majority rule based on many votes, it naturally requires tremendous cost, time, and human efforts.  Thus, our challenge lies in determining whether or not we can assist the Tribunal by machine learning.  Although toxic behavior has been known as hard to define, we have a huge volume of labeled data. 

In this paper, we explore toxic behavior using supervised learning techniques.  We collect over 10 million user reports involved in 1.46 million toxic players and corresponding crowdsourced decisions.  We extract 534 features from in-game performance, user reports, and chats.  We use a Random Forest classifier for crowdsourced decision prediction.  Our results show good performance in detecting overwhelmingly majority cases and predicting crowdsourced decisions on them.  In addition, we reveal important features in predicting crowdsourced decisions.  We demonstrate good portability of our classifier across regions.  Finally, we estimate the practical implications of our approach, potential cost savings, and victim protection.

We make several contributions.
To the best of our knowledge, we provide the first demonstration that toxic behavior, notoriously difficult to objectively define, can be detected by a supervised classifier trained with large-scale crowdsourced data. 
Next, we show that we can successfully discriminate behavior that is overwhelmingly deemed innocent by human reviewers.
We also identify several factors which are used for the decision making in the Tribunal.
We show that psychological measures of linguistic ``goodness'' successfully quantify the tone of chats in online gaming.
Finally, we apply our models to different regions of the world, which have different characteristics in how toxic behavior is exhibited and responded to, finding common elements across cultures.

In Section~\ref{sec:background} we give background on LoL and its crowdsourced solution for dealing with toxic behavior.
In Section~\ref{sec:related} we review previous literature.
In Section~\ref{sec:data_collection} we provide details on our dataset.
In Section~\ref{sec:research_goal} we outline our research goals.
In Section~\ref{sec:features} we explain features extracted from in-game performance, user reports, and linguistic signatures based on chat logs.
In Section~\ref{sec:models} we build several models using these features.
In Section~\ref{sec:results} we propose a supervised learner and present test results.
In Section~\ref{sec:discussion}, we discuss practical implications and future research directions, and we conclude in Section~\ref{sec:summary}. 

\section{Background}\label{sec:background}

To help readers who are unfamiliar with LoL, we briefly introduce its basic gameplay mechanisms, toxic playing in LoL, and the Tribunal system. 

\subsection{League of Legends}

LoL is a match-based team competition game.
Teams are most often composed of five players who are randomly matched together, and friends can also form pre-made teams.

LoL features a symmetric map with three paths, or ``lanes''.
The lanes are colloquially known as ``top'', ``mid'', and ``bot'' and have a ``jungle'' area between them.
The goal of LoL is to penetrate the enemy team's central base, or ``Nexus'', and destroy it.
To destroy the Nexus, players must first destroy towers in at least one lane.
Typically, when a match starts players choose to go to one particular lane, often taking on different roles depending on the character they chose to play.

Players' in-game characters are called champions.
Riot has released 115 champions as of September 2013.
A weekly rotation of 10 champions is offered to all players, but they can also be permanently purchased via points earned through play or real world money.
Each champion has different strengths and thus naturally lend themselves to the different styles of play expected from a particular lane.

\subsection{Reporting Toxic Players}

After a match, players can report toxic players in one of 10 predefined categories: assisting enemy team, intentional feeding, offensive language, verbal abuse, negative attitude, inappropriate $[$handle$]$ name, spamming, unskilled player, refusing to communicate with team, and leaving the game$/$AFK $[$away from keyboard$]$.  

Intentional feeding indicates that a player allowed themselves to be killed by the opposing team on purpose, and leaving the game is doing nothing but staying at the base for the entire game. 
To understand the motivation behind intentional feeding and assisting the enemy team, we present a common scenario when such toxic play happens.
LoL matches do not have a time limit, instead continuing until one team's Nexus is destroyed.
If a team is at a disadvantage, real or perceived, at some point during the match, they may give up by voting to surrender.
However, surrendering is allowed only after 20 minutes have passed, and the vote requires at least four affirmatives to pass.
If a surrender vote fails, players who voted for the surrender may lose interest and regard continuing the game as wasting time.
Some of these players exhibit extreme behavior in response to the failed vote, such as leaving the game$/$AFK, intentional feeding, or assisting the enemy.
Leaving the game is particularly painful since being down even one player greatly reduces the chance of a ``come from behind'' victory.
In other online games, leaving is not typically categorized as a toxic because it usually does not harm other players in such a catastrophic manner.

\subsection{LoL Tribunal}

To take the subjective perception of toxic behavior into account, Riot developed a crowdsourcing system to judge whether reported players should be punished, called the \textit{Tribunal}.
The verdict is determined by majority votes.

When a player is reported hundreds of times over dozens of matches, the reported player is brought to the Tribunal\footnote{http://tinyurl.com/stfunub4}.  
Riot randomly selects at most five reported matches\footnote{http://tinyurl.com/stfunub5} and aggregates them as a single case.
In other words, one \emph{case} includes up to 5 \emph{matches}.  
Cases include detailed information for each match, such as the result of the match, reason and comments from reports, the entire chat log during the match, and the scoreboard, that reviewers use to decide whether that toxic player should be pardoned or punished.  
To protect privacy, players' handles are removed in the Tribunal, and no information about social connections are available.  
We note that our dataset does not include reports of unskilled player, refusing to communicate with team, and leaving the game in our datasets, even though players are able to choose from the full set of 10 predefined categories\footnote{To the best of our knowledge, this is intentional on the part of Riot.}.

The Tribunal also institutes mechanisms to maintain the quality of reviewers' tasks.
One is a limit on the number of cases that reviewers can see a day.  
Next is a minimum duration for decision making, limiting mechanical clicks without careful consideration.
Another is a skip feature for difficult cases. 
 
After some time has passed, reviewers can see the final crowdsourced decisions for the cases they reviewed as well as the level of agreement (majority, strong majority, or overwhelming majority).
To encourage user participation, Riot adopts reviewers' accuracy score and ranking as gamification element.

\section{Related Work}\label{sec:related}

\subsection{Toxic Playing in Online Games}

Toxic behavior in online games is a form of cyberbullying, defined as repetitive intentional behavior to harm others through electronic channels~\cite{Smith08}.
Computer-mediated-communication through electronic channels without face-to-face communication lacks social psychological influences and can lead to more hostile interaction~\cite{ho2008social}.
In psychology and education, offline bullying has been actively researched~\cite{Olweus96}, and offers a theoretical background to understand cyberbullying.  

Griefing is a term describing cyberbullying in online gaming, and those who enjoy the act of disrupting other players are called grief players (``griefers'')~\cite{Foo04,McKenna00}.  
Griefers make other players annoyed and feel fatigued.
Sometimes victims even leave the game~\cite{Mulligan03}, exhibiting toxic behavior themselves to escape beratement.

Although griefing is intuitively understood, its boundary is unclear~\cite{Chesney09,Foo04,Lin05,Mulligan03} because customs, rules, or ethics can be different across games~\cite{Warner05}. 
In addition, the perception of grief behavior is unique across individuals.  
As a consequence, even griefers themselves may not recognize what they have done as griefing~\cite{Lin05}.
This inherent vagueness makes it hard to understand grief behavior. 

A few studies have characterized grief playing.  
Foo and Koivisto divide grief behavior into four categories: harassment, power imposition, scamming, and greed play~\cite{Foo04}.  
They focus on the intention of behavior and distinguish griefing from \emph{greed playing} because motivation behind greed playing is usually for the win instead of disrupting other players.  
Barnett discovers that griefing is one of factors provoking anger in World of Warcraft by survey of 33 participants~\cite{Barnett10}.
Chen et al. correlate personality and grief playing.
They reveal that players who enjoy the anonymous experience tend to like the motivation of grief playing, such as ``I like the experience of wanting to feel powerful while playing online games''~\cite{Chen09}.

\subsection{Challenges in Crowdsourcing}
 
While crowdsourcing, coined by Jeff Howe~\cite{howe2006rise}, has had huge impacts on a variety of areas, such as answering queries~\cite{franklin2011crowddb}, assessing visual design~\cite{heer2010crowdsourcing}, translating texts~\cite{zaidan2011crowdsourcing}, conducting user studies~\cite{kittur2008crowdsourcing}, and collecting disaster responses~\cite{goodchild2010crowdsourcing}, inherent noise in outcomes by laborers has received much attention~\cite{ipeirotis2010quality,paolacci2010running}.
This issue is not limited to amateur labor groups.
Voorhees demonstrate that even experts show low levels of agreement on subjective problems, such as judging the relevance of documents to a given topic~\cite{voorhees2000variations}.

Majority votes or overlapping labels are popular methods to improve the overall quality of error-prone labelers.  
Nowak and R{\"u}ger demonstrate that majority voting on crowdsourcing is able to filter noisy judgments and to improve the quality of experts' annotations~\cite{nowak2010how}.
Sheng et al. investigate how overlapping labels for low agreement tasks improve quality, especially when labelers are imperfect~\cite{sheng2008get}.
They find that a selective repeated-labeling strategy can improve data quality substantially.
Some studies benefit from maximizing the utility of more expert workers.
Snow et al. propose giving different weights to each labeler based on how their accuracies improve the label quality~\cite{snow2008cheap}.  

Crowdsourcing has been widely used for producing ground-truth labels for data-driven machine learning, such as labeling sentiment in social media~\cite{brew2010using}.
To obtain more accurate classifiers from crowdsourced labels with noise, Brew et al. report that a classifier with training sets only from labelers showing high agreement achieves better accuracy than with less agreement~\cite{brew2010using}.
Also, Kumar and Lease show that incorporating the accuracies of each labeler has a large impact on the overall accuracy of a supervised learner~\cite{kumar2011modeling}.
Tang and Lease propose semi-supervised learning with many unlabeled tasks but a small set of expert-labeled tasks~\cite{tang2011semi}.  

Our approach, which regards supervised learning and crowdsourcing as complementary mechanisms to balance loads, is inspired by two recent systems: CrowdFlow and CrowdSynth.
Quinn et al. develop CrowdFlow~\cite{quinn2010crowdflow}, an integrated framework of human efforts and machine learning classifiers, and discuss the efficiency of the crowd-assisted machine learning~\cite{quinn2011human}.
As a sample application, they first translate texts via Google Translate, and then let experts identify problematic parts and offer alternatives.
They do not automatically identify difficult parts of texts but require manual labor. 
Kamar et al. present a prototype of CrowdSynth~\cite{kamar2012combining} to guide workers with predicting the labels in a well-known citizen science application called Galaxy Zoo.
Their models are derived from relatively rich data, such as detailed records for labels, workers, and combinations of them.

\section{Data Collection}
\label{sec:data_collection}

We developed distributed crawlers to collect over 1.46 million cases in the Tribunal, made up of 10 million user reports.
We carefully designed our crawlers not to degrade the performance of Riot's web servers.
In addition to the parsing time of the crawled contents, we explicitly let our crawlers be idle for seconds before the next request.

Riot divides the world into several regions, each served by dedicated servers due to quality of service concerns.
We collected data from North America (NA), Western Europe (EUW), and South Korea (KR) servers by considering the representativeness of cultural uniqueness and the familiarity of authors.  
We reasonably assume most of the players connect to the servers in the region closest to them for the best user experience.

As a result, in April 2013, we crawled all available cases from the Tribunal of the each of the three regions.  We summarize our data collection in Table~\ref{tbl:data_collection}.
We reiterate that a case includes multiple matches, which in turn contain one or more reports.
\begin{table}[ht!b]
\begin{tabular}{ c r r r r }
	\hline
		& \multicolumn{1}{c}{EUW} & \multicolumn{1}{c}{NA} & \multicolumn{1}{c}{KR} & \multicolumn{1}{c}{Total} \\
	\hline
	Cases & 649,419 & 590,311 & 220,614 & 1,460,344 \\
	Matches & 2,841,906 & 2,107,522 & 1,066,618 & 6,016,046 \\
	Reports & 5,559,968 & 3,441,557 & 1,898,433 & 10,899,958 \\
	\hline
\end{tabular}
\caption{Data collection summary}
\label{tbl:data_collection}
\end{table}

\section{Research Questions}
\label{sec:research_goal}

With our large-scale labeled data via crowdsourcing, we raise two research questions.  

\vspace{2mm}
RQ1: Can we predict the crowdsourced decisions?
 
Since the perception of toxic behavior is subjective and different across individuals, the Tribunal deals with toxic behavior by a majority rule based on crowdsourcing.
It has worked quite well, but requires a long time to obtain enough votes.
Our research question is to validate whether a certain part of the Tribunal can be assisted by machine learning.  
A few considerations are carefully addressed here.

\textbf{Defining machine learning tasks} ~~~  We can define various machine learning tasks on crowdsourced decisions in the Tribunal.
Classifying 6 different combinations of decision and level of agreements, dividing cases into punished or pardoned, extracting high agreement cases, and recognizing less agreement cases are all possible tasks that machine learning could help.
It is not only about the accuracy of the classifier but also about the application. 

\textbf{Refining training set} ~~~ A Tribunal decision is either punish or pardon with a level of agreement: majority, strong majority, and overwhelming majority.
Previous literature demonstrate that crowdsourced responses with high agreement are better for training a classifier than less agreement~\cite{brew2010using}.
Unfortunately, it is unknown how Riot divides these levels of agreement.
We thus create different training sets and compare the accuracy of trained classifiers.

\textbf{Using non-linguistic features only} ~~~  LoL has a global userbase in a few tens of countries across the world.
This implies that chat logs in the Tribunal are not always written in English.
For example, most messages in the Korean Tribunal are written in Korean.
Various languages can potentially limit the portability of the classifier if it largely depends on the textual information left in chats.
In other words, more non-linguistic features increase the generality of the classifier and bring higher practical impacts.

\textbf{Detecting sentiments in chats} ~~~ This is the inverse of the above, maximizing the benefit of linguistic features.
Many methods have been developed for detecting sentiments conveyed in texts.
Our intuition is that negative sentiments might be captured from toxic behavior or other players' reaction when toxic playing occurs.

\vspace{2mm}
RQ2: What do the important features imply?

The next research goal is understanding decision-making in the Tribunal from the important features observed through supervised learning.  
Which features are important for predicting crowdsourced decisions?
What do they mean?
Answering these questions leads to several interesting challenges.  
First of all, features we find could be a huge advance in online violence research.
Toxic behavior has been typically considered hard to define.
If we obtain a good quality supervised-learned classifier, it indicates the important building blocks in defining and understanding toxic behavior.
Then, we draw a subsequent question.  
Can we apply our classifier or the important features to other games, or Internet communities?
If important features in our classifier are LoL-independent, the answer to the question will be in the affirmative.

\section{Features}\label{sec:features}

The Tribunal can be seen as a 2-stage crowdsourced solution.
Stage 1 is the per-match reporting done by players that actually experienced the alleged toxic behavior.
Stage 2 is the Tribunal case-by-case judgments.
Toxicity is not having a bad game (possibly perceived as feeding or assisting the enemy) or having a bad day and lashing out at a teammate (harassment).
According to Riot, a certain threshold of reports from stage 1 must be met before moving on to stage 2, which reveals \emph{repeated} toxic behavior.
The reason that stage 2 is necessary has to do with the vagueness of toxic behavior.
Consider a player who is not very skilled.
This player might exhibit tendencies that could be interpreted as toxic, for example intentionally feeding.
This is compounded by attribution theory where a negative outcome (losing a match) triggers a search for an external cause.
I.e., when a team loses, as is the the case in the majority of reported matches~\cite{chisubmission}, reporters might attribute the loss to a poorly performing player.
Although there is an unskilled player report type, players are aware that no punishment is handed out for this category and thus might choose one of the punishable offenses instead.
Thus, the second stage removes the subjectivity associated with direct interaction with the possibly toxic player, as well as providing a more complete view of the accused's behavior over multiple matches.
Players that have invested significant time, and are thus familiar with LoL, are able to pick out patterns of toxic behavior when reported matches are combined into a case. 

The challenge lies in representing the parsimonious data presented to Tribunal reviewers in a form digestible by machine learning algorithms.
We thus extract summarized statistics from each Tribunal case.
We make use of two primary sources of information: 1)~domain specific values extracted from the results of reported matches, and 2)~the information provided by the stage 1 Tribunal participants.

There are, unfortunately, several points of variation when it comes to extracting the in-game values.
First, each case has a varying amount of matches with no guarantee on the sequence in which the matches took place.
Second, because of the variety of game play in LoL, there is no guarantee that matches are directly comparable, especially across different players.
For example, a player with below average skill is likely to have a lower KDA (an in-game performance metric explained in the next section) than a player with higher skill.
Now assume that the low skill player is not toxic, while the high skill player \textit{is} toxic, yet both are reported for intentional feeding.
There might not be a way of discriminating between the two using just KDA.

Although we average the per-match statistics across all matches with a given report type for each case, we could also use the worst/best matches.
This is somewhat problematic as it requires a definition of worst and best.
We include the standard deviation of each averaged statistic as a feature, which provides a sense of inter-match performance differences.

We then augment each instance with information provided by the Stage 1 Tribunal participants.
Namely, we make use of the number of allied and enemy reports in a given match, the number of reports where the reporter included optional human readable text about the offense, and the most common type of behavior reported for a given match.
For each possible report type, we compute the relevant statistics across all matches in the case with said most common report type.

\subsection{In-game Performance}

In-game performance is the category of features that most requires the input of \emph{experts}.
LoL is a complicated game and the meaning of the various match-related statistics is unlikely to be divined by a reviewer, especially with respect to toxic behavior, without having investing a significant number of hours in gameplay themselves.
Nevertheless, they are the richest and easiest features to represent to a computer and so we extract a set of relevant statistics from the matches in each Tribunal case.

We first begin with the most basic statistic, one that is common to nearly all competitive games: kills, deaths, and assists.
Kills and deaths are relatively self-explanatory: simple counts of the number of enemies a player killed and the number of times said player died.
Likewise, an assist is awarded to a player that participated in eliminating an enemy, but did not land the killing blow.
The details of what qualifies an assist varies per game, but LoL awards an assist to any player that did damage or contributed passively (e.g., healed a teammate that landed the killing blow) within 10 seconds prior to the death of an enemy.
Kills, deaths, and assist are raw scores, but are often normalized to a KDA metric, defined as:
\[
KDA = \frac{kills + assists}{deaths + 1}
\].

Unfortunately, due to the reliance on teamwork in games like LoL, a single toxic player can severely impact his teammates abilities to perform at a high level.
For example, an intentional feeder is supplying the enemy team with gold and experience points, allowing them to acquire powerful items and abilities much faster than the feeder's allies.
In turn, this can result in a low KDA not only for the toxic player, but makes it difficult for his allies to maintain a good KDA as well.
For this reason, it might be difficult to distinguish toxic players based solely on KDA and our initial analysis~\cite{chisubmission} indicated reviewers were not basing decisions only on KDA.
However, two other statistics, damage dealt and received, might shed additional light on toxic players.

In LoL, attacks do a certain amount of base damage to other players, removing a portion of their hit points (``life'').
When a player's hit points reach 0, he dies, a kill (with associated gold and experience) is awarded to his killers, and he must wait a certain period of time to ``respawn'' and return to the fight.
The base damage is modified depending on various items that players purchase during the match.
Anecdotally, toxic players in the feeding and assisting enemy categories will not buy items that aid in offense or defense.  
Thus, we might expect a feeder to have very low damage dealt and very high damage received relative to his teammates who have made purchases of useful items; even though they might not have the power to actually kill enemies (due to a gold and experience advantage given to the other team by the feeder), fair players' efforts are likely to show in terms of damage.
Seeing which items a player bought could give more details, but it is overly specific and loses generality.

Next, we include the total gold and gold per minute earned by the offender.
In LoL players earn gold in several ways: 1)~passively at a pre-determined rate, 2)~destroying towers, 3)~kills or assists, and 4)~killing creeps, which are computer controlled monsters.
Gold is known to be a primary determinant of a match's outcome.

The final performance related feature is time played, which is useful for detecting leavers or players going AFK.
In total, there are 364 features generated from in-game performance values only.
This number is large because we group per-match values based on the most common report type.

\subsection{User reports}

The first stage of the Tribunal, reports submitted players who directly observed the behavior, provides us with several pieces of information.
First, we know the most common category of behavior reported per match.
Next, we know how many allies and enemies made a report.
Finally, reports can include a short (500 character limit) comment from the reporter.
Intuitively, the extra effort required to add a comment to a report might indicate the intensity of the toxic behavior.
Again, we group the per match values in the case based on the common report type for that match, resulting in a total of 28 features.

\subsection{Chats}

\begin{figure}[ht!b]
	\includegraphics[scale=1.0]{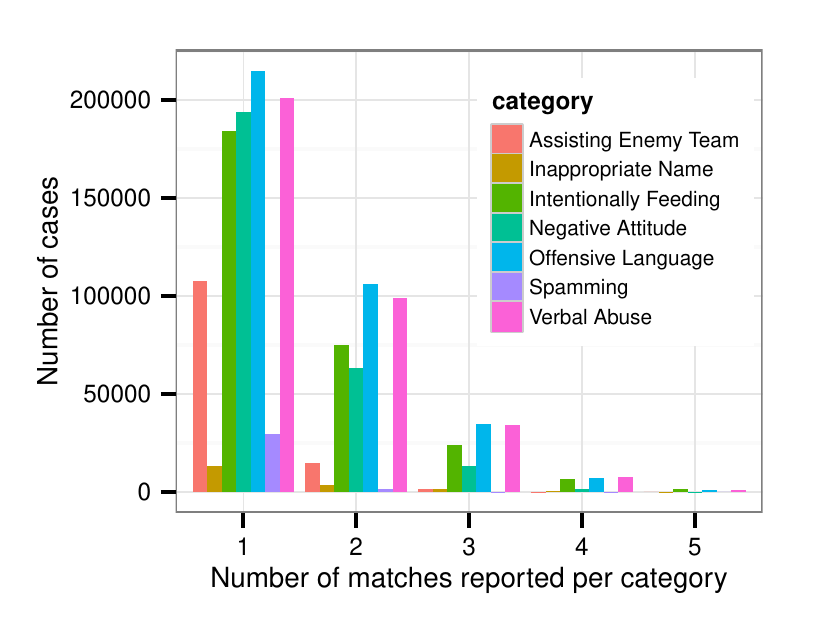}
	\caption{The number of matches reported in a case for each category of toxic behavior per Tribunal case.}
	\label{fig:report-categories}
\end{figure}

As seen in Figure~\ref{fig:report-categories}, around 60\% of cases have more than about 25\% of the matches in them reported for offensive language or verbal abuse. 
This indicates that the observed toxic behavior was expressed (at least partially) via the chat system.
For this reason, we intuit that the chat logs have predictive power.
Linguistic analysis is an area of intense research~\cite{frank2013happiness,golder2011diurnal,tumasjan2010predicting}, and the corpus of chat logs in our data set would provide an interesting case for cutting edge techniques.
We use an intuitive and straightforward method, ``happiness'' index.

\begin{figure*}
	\includegraphics[scale=0.24]{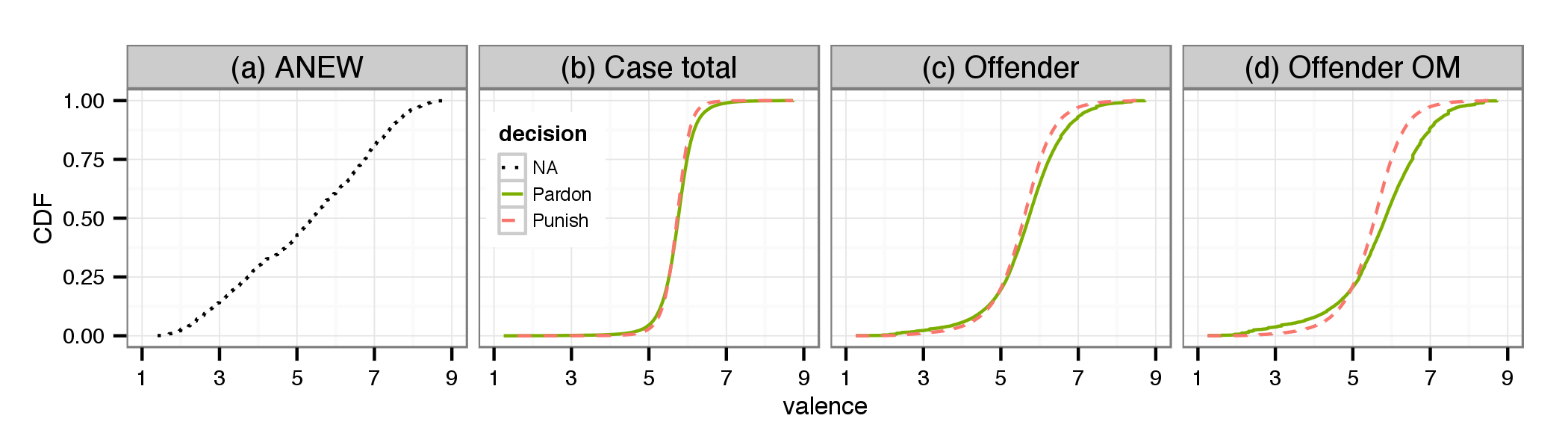}
	\caption{CDF of valence scores.}
	\label{fig:valence-scores-row}
\end{figure*}

Happiness, and its relation to language, is a widely studied area of psychology~\cite{bellezza1986words}.
For example, in the Affective Norms for English Words (ANEW) study~\cite{bradley1999affective}, participants graded a set of 1,034 words on a \emph{valence} scale of 1 to 9 (in 0.5 increments).
Valence is the psychological term for the attractiveness (positive) or aversion (negative) to something; in this case a word.
In other words, valence quantifies the ``goodness'' or ``badness'' of a word.
Valence scores in the ANEW dataset are well distributed, as can be seen in Figure~\ref{fig:valence-scores-row} (a), which plots the distribution of scores for all words in the ANEW dataset.

The ANEW study polled both female and male respondents.
Riot reports that 90\% of LoL players are male\footnote{http://tinyurl.com/stfunub6}, and we thus use male respondent scores only.
Although gender swapping often occurs in social games~\cite{lou2013gender}, according to Flurry's report\footnote{http://tinyurl.com/stfunub7}, action and strategy are the top two genres that females usually avoid.  Also, positive effects of gender swapping, enjoying a kind of ``second'' life, do not apply to LoL.  

As demonstrated by Dodds and Danforth~\cite{dodds2010anew}, valence scores for individual words can be used to estimate the valence for a larger corpus of text.
The valence of a piece of text is defined as:
\[
v_{text} = \frac{\sum_{i=1}^{n} v_i f_i}{\sum_{i=1}^{n}f_i}
\],
where $v_i$ is the valence score of the $i$th word from the ANEW study, and $f_i$ is the number of times said word appears in a given piece of text.

While we acknowledge that chat logs are likely to contain typos and abbreviations, $v_{text}$ has been shown to be robust across genres of text, including tweets~\cite{dodds2010anew,mitchell2013geography}, another medium where we might expect ``Internet-style'' speech patterns.
For cases where no ANEW words were present, we define $v_{text} = 0$.

Figure~\ref{fig:valence-scores-row} (b) plots the distribution of valence scores of all messages sent in a case for both pardoned and punished cases where $v_{text} \geq 1$.
A two-sample Kolmogorov-Smirnov test confirms the distributions are different.
When compared to Figure~\ref{fig:valence-scores-row} (a), we see that ``verbal'' communication in LoL is generally neutral: most valence scores fall between 5 and 6.
Further, cases that resulted in a punishment tend to have a lower valence score when compared to pardoned cases.
This indicates that the chat logs are likely to be valuable in detecting toxic behavior.

When looking at the scores of messages sent only by potential toxic players, offenders, in Figure~\ref{fig:valence-scores-row} (c), it becomes clear that toxicity is present in a quantifiable way within the chat logs.
The distributions for both punished and pardoned offenders is lower than the valence of the entire chat logs.
The mean for punished and pardoned users are 5.725 and 5.779, respectively.
Pardoned users are indeed more likely to have higher valence scores.  
Interestingly, the difference is mainly present in terms of ``above average'' ($\geq 5$) valence scores for pardoned users as opposed to a tendency towards below average scores for punished players.
We also discover that the difference between punished and pardoned offender becomes bigger if more reviewers are agreed.
Figure~\ref{fig:valence-scores-row} (d) shows the valence score of toxic players when overwhelming majority is agreed.
The mean for only those who are punished or pardoned in this case are 5.699 and 5.751, respectively.

In addition to the scores described above, we also include the valence scores for bystanders and victims for each report category.
Here, we treat verbal abuse, offensive language, and negative attitude differently from the other categories.
For these cases we have previously observed that reports by enemies are much more common.
This can be attributed to bystander theory, which says that bystanders, i.e., those not directly harmed by bad behavior, are much less likely to take action against it.
In the case of, e.g., intentional feeding, not only are enemy teams not directly harmed by the behavior, they actually receive a benefit.
While the quality of the competition may decrease, the odds of a win are much more in their favor.

When it comes to chat based offenses, however, a toxic player can lash out at everyone in the match.
He can insult the enemy team when they are performing well, and trash talk when they are performing poorly.
For example, a common insult in LoL is to call someone a ``noob,'' slang for ``newbie,'' implying a lack of ability.
It is a catch-all negative term used as a response to criticism, to call out poor play, as a form of trash talk, and just plain meanness.
The word ``noob'' appears 3,551,328 times in the chat logs for the NA region; an average of 6 times per case and 1.7 times per match, with players under review saying ``noob'' at more than double the rate per message sent as non-offenders.
Because these communication based categories can affect both enemies of the offender, allies, or neither, we consider their victims to be the offender's allies if only allies reported him, enemies if only enemies reported, and all players if both enemies and allies reported him.

With the above in mind, we extract 60 features per case from the chat logs.

\section{Models}\label{sec:models}

As we introduced above, we extract features from different categories.
We then build separate models for each category, and a full model to contain all the features: an in-game performance model, a user report model, a chat model, and a full model.
The intuition behind these basic models is comparing different sources.

First, in the in-game performance model, we can divide features by offender performance and other players' performance.  
Offender performance can reflect intentional feeding.
Players with noticeably bad performance might die over and over intentionally.
We note that non-toxic \emph{unintentional} feeding does occur, but only \emph{intentional} feeding is toxic behavior.
Since there is an unskilled player Stage 1 report category that is unused in the Tribunal, we ensure that our classifier is trained only on reviewers' inference of intention, and not a judgment of player skill.
Other players' performance relates to the team competition aspect of LoL.  
Attribution theory says that individuals will look for external causes of failure~\cite{weiner1980cognitive}.
The most obvious cause would be poor performance by an ally, and is likely to manifest as verbal abuse (harassment).
In other words, a toxic player might lash out at the worst performing ally due to the perception that a loss was the fault of said ally.
We hypothesize that the intensity of the verbal abuse, and thus the likelihood of punishment in the Tribunal, increases as the offender's performance diverges from the worst performing player on his team.

A less intuitive reasoning in favor of this model is that a poor performance by a player \textit{does} have an impact on the rest of his team.
Our previous analysis indicates that KDA alone is insufficient in predicting the judgment of the tribunal: there was no correlation with KDA and corresponding Tribunal decision~\cite{chisubmission}.
I.e., players with the best KDA were about as likely to be punished as those with the worst.

The user report model depends on how players in a match perceive toxic playing.
Although user perception on toxic behavior is different, more reports in a single match means more people recognize it, and implies more severe toxic behavior.
In our initial analysis~\cite{chisubmission}, we find that the number of reports in a single match is highly correlated with the likelihood of being punished.

The chat model deals with behavior that is expressed \emph{only} via communication; verbal abuse and offensive language.
Additionally, Tribunal reviewers can see the final in-game performance but not how the match played out over time. 
Chats are thus the only source to give reviewers \emph{context} about what happened and what other players thought about it.

\section{Results}\label{sec:results}

\subsection{Decision confidence and classifier training}

We first answer whether or not reviewer agreement is relevant when training a classifier.
To do this, we grow forests from only cases of a given agreement.
We then evaluate each classifier with a separate test set for each agreement.
The intuition is that the cases with the highest level of agreement from the reviewers display the most egregious toxic behavior, providing a baseline for the remainder of cases.

\begin{figure*}[t!]
\begin{center}
\begin{minipage}{55mm}
\begin{center}
\includegraphics[width=55mm]{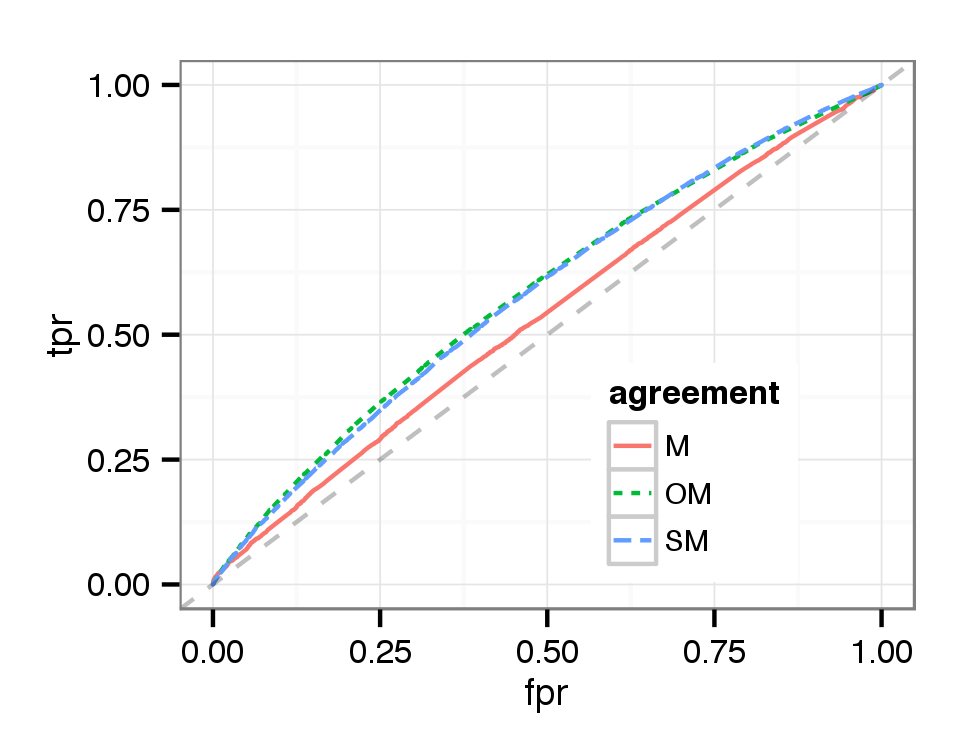}\\
\end{center}
\centerline{(a) Majority}
\end{minipage}
\begin{minipage}{55mm}
\begin{center}
\includegraphics[width=55mm]{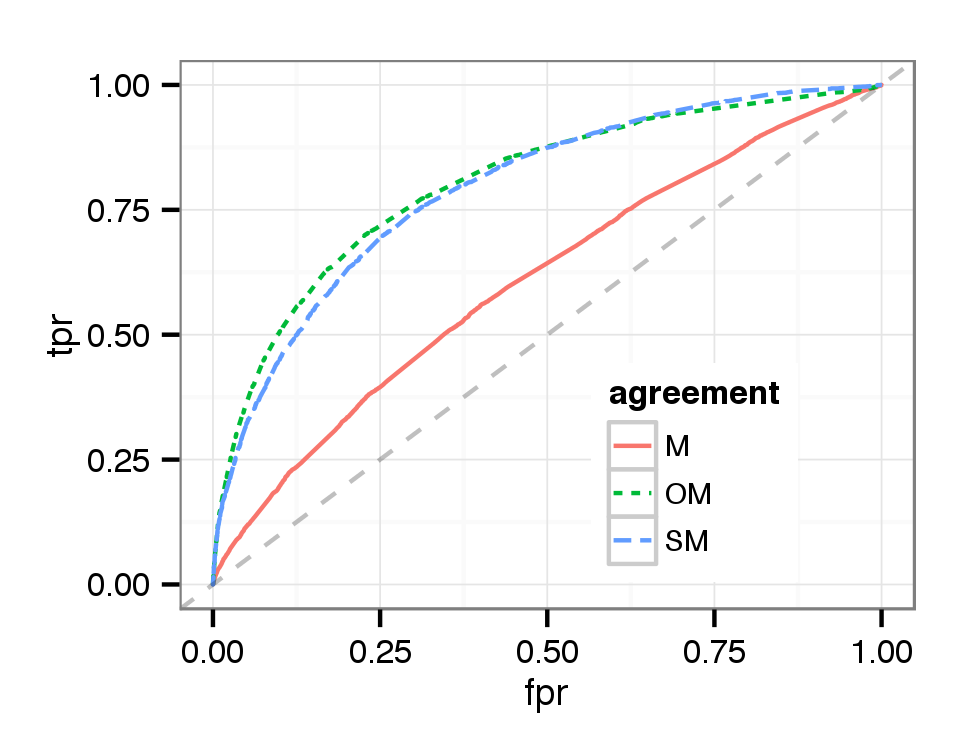}\\
\end{center}
\centerline{(b) Strong Majority}
\end{minipage}
\begin{minipage}{55mm}
\begin{center}
\includegraphics[width=55mm]{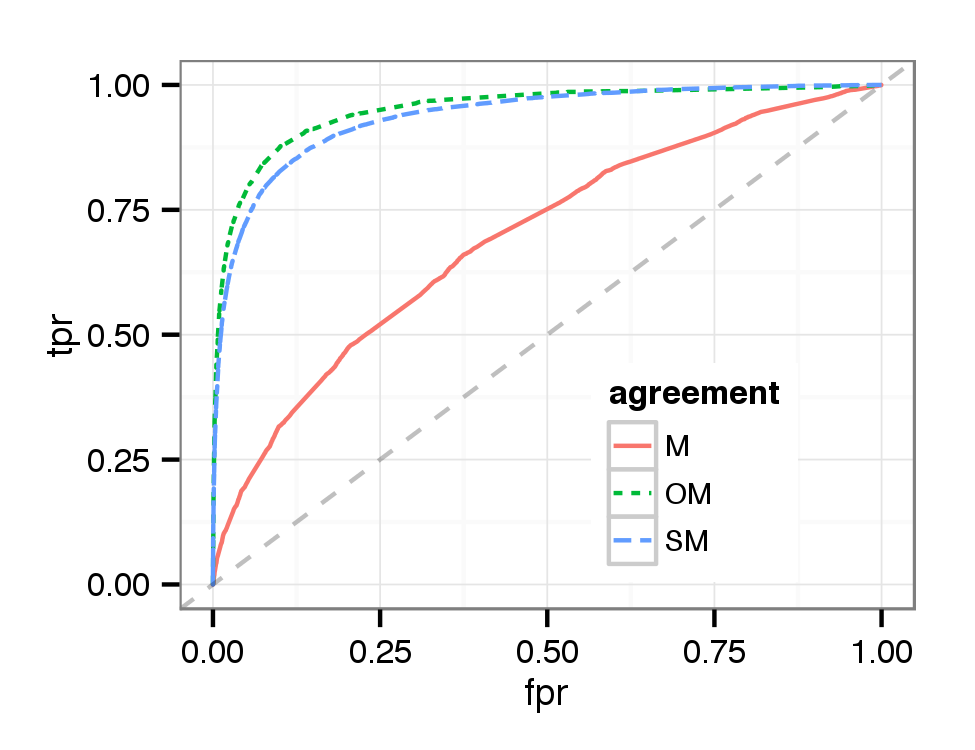}\\
\end{center}
\centerline{(c) Overwhelming Majority}
\end{minipage}
\caption{[Best viewed in color.] ROC curves for cases with a majority / strong majority / overwhelming majority decision using classifier trained from majority cases (``M''), strong majority cases (``SM''), and overwhelming majority cases (``OM'')}
\label{fig:majority-roc}
\end{center}
\end{figure*}

Figure~\ref{fig:majority-roc} plots the ROC curves for testing sets of each agreement type with classifiers trained from varying agreement cases.
We observe that training the classifier with overwhelming majority decisions results in the highest AUC across all cases.
Our ability to distinguish between guilty and not guilty increases with the level of agreement that we train the classifier with.
This is consistent with previous research~\cite{brew2010using}.
While the classifier trained with overwhelming majority is the most discriminating across the board, training with strong majority cases has similar performance, while performance drops off considerably when training with the majority decision cases.

This experiment has several implications.
First, it might be beneficial to look at the extremes of behavior, clear cut pardons and punishes, to better predict borderline cases.
Second, it might be fruitful to predict decisions based on confidence.
I.e., finding the most obviously guilty or innocent individuals, leaving the borderline cases to human reviewers.
Third, it reveals the difficulty in discriminating between all but the most egregious offenses.

\subsection{What are Tribunal decisions based on?}

We now address what information Tribunal reviewers might be basing their decision on.
We present a few learning results to show the performance of our Random Forest classifier.  

\begin{figure}[t]
	\centering
	\includegraphics[width=\columnwidth]{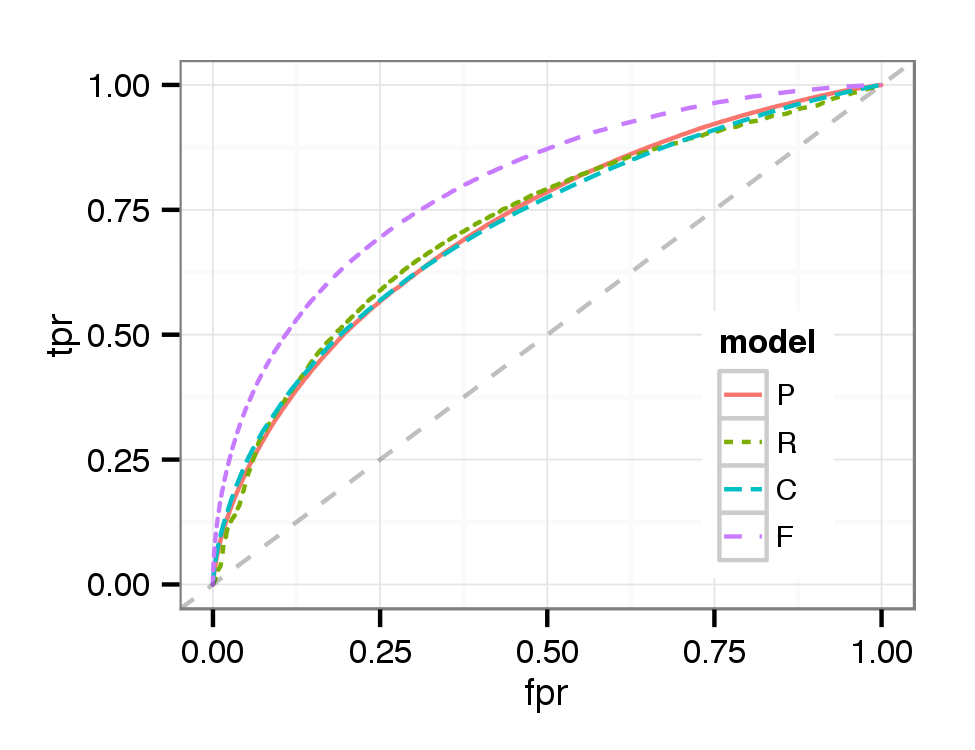}
	\caption{[Best viewed in color.] ROC curve for predicting Tribunal decisions with models using in-game performance (P), user report (R), chats (C), and all available features (F).}
	\label{fig:offender-allies-enemies-roc}
\end{figure}

We begin with the performance of predicting decisions, pardon or punish without considering the agreement level.
Figure~\ref{fig:offender-allies-enemies-roc} presents ROC curve for predicting decisions, punish or pardon, by different models.  
AUCs are 0.7187, 0.7195, 0.7157, and 0.7991 for the performance, report, chat, and full models, respectively.
We observe that each model shows comparable performance. 

Table~\ref{tbl:feature-ranks} shows the 5 most important variables for predicting decisions.
Due to lack of space, we omit the important variables for each category of toxic behavior, but it is similar across the categories.

\begin{table}[t]
        \centering
        \begin{tabular}{|r|l|}
        \hline
        Rank & Feature \\
        \hline
        \multicolumn{2}{|l|}{Performance only} \\
        \hline
        1 & verbal.abuse.enemies.kda \\
        2 & verbal.abuse.enemies.gpm \\
        3 & verbal.abuse.offender.deaths \\
        4 & verbal.abuse.enemies.kda.avg.per.player \\
        5 & verbal.abuse.offender.kda \\
        \hline
        \multicolumn{2}{|l|}{Report only} \\
        \hline
        1 & verbal.abuse.allied.report.count \\
        2 & verbal.abuse.allied.report.comment.count \\
        3 & intentionally.feeding.allied.report.count \\
        4 & intentionally.feeding.allied.report.comment.count \\
        5 & offensive.language.allied.report.count \\
        \hline
        \multicolumn{2}{|l|}{Chat only} \\
        \hline
        1 & case.offender.valence \\
        2 & verbal.abuse.offender.chat.msgs \\
        3 & offensive.language.offender.chat.msgs \\
        4 & verbal.abuse.offender.valence \\
        5 & verbal.abuse.total.chat.msgs \\
        \hline
        \multicolumn{2}{|l|}{Full} \\
        \hline
        1 & case.offender.valence \\
        2 & verbal.abuse.allied.report.count \\
        3 & verbal.abuse.offender.chat.msgs \\
        4 & offensive.language.offender.chat.msgs \\
        5 & verbal.abuse.allied.report.comment.count \\
        \hline
        \end{tabular}
        \caption{The top 5 ranked features from an information gain evaluator for Tribunal decisions.}
        \label{tbl:feature-ranks}
\end{table}

In the performance model, we find that enemy performance is a good predictor for decisions because offender or ally performance is relative in team competition games.  
Also, offender performance itself is important for decision making in the Tribunal.
Interestingly, the number of deaths is more important than KDA.
This might relate to partner blaming.
A toxic player might blame teammates, e.g., saying ``noob'', when he dies.  
The implication is the toxic player died because allies did not help him.  
In this situation, the number of deaths could be a good measure to reveal that the toxic player performed poorly rather than his allies not helping out.

In the user report model, top variables are related to how many reports are submitted.
The more reports, the more likely the Tribunal will decide to punish.
This strongly agrees with our previous analysis~\cite{chisubmission}.
Additionally, we find that reviewers care about short comments in user reports.
This implies that a user interface encouraging comments might be helpful for crowdsourced decision-making.

In the chat model, we find that the most important variable is the valence score of offender no matter what the reported toxic category is.  
This agrees with our intuition that reviewers can see the context from chats and infer what happened.  They gain insights from chat not only for verbal abuse or offensive language, but also other categories of toxic behavior.  
Also, this demonstrates that the ANEW dataset works well even with chats in online games. 
The second and third most important variable are the number of messages sent by the offender when the reported reason is verbal abuse and offensive language.
This implies that the likelihood of toxic behavior goes up if the offender talks more.
This can be easily recognized in real-time and thus pragmatically used for warning toxic players through visual cues.
The fourth important variable is the valence score when the behavior is reported as verbal abuse.
This is straightforward to understand.  
We find that number of total messages sent when the report reason is verbal abuse is the fifth important variable. 
This is the only feature in the top five that is not only related to the offender, but also others.
If more players are involved in a quarrel, it is a strong sign of verbal abuse having occurred.

The top 5 features in the full model are entirely from the chat and report models.
The total valence of the case is the number one feature, which highlights how much toxic behavior is visible/expressed via in-game communication.
The second most important feature in the full model comes from the report only model, highlighting how our approach dovetails with the first crowdsourced stage of the Tribunal.
The hints provided by those that directly experience toxic behavior are useful not only to human reviewers, but, for an algorithmic solution as well.
Next, we note that the 6th and 7th most important features in the full model are from the performance model.
Thus, while in-game performance numbers \emph{are} a predictor of toxic behavior, context is key.

We also look into the top 5 important variables in predicting overwhelming majority pardon and punish, respectively.
Due to lack of space, we omit the details but we highlight two findings by comparing them.  
One is that there are great discrepancies of important variables between the model for predicting overwhelming majority pardon and punish.  
It implies that reviewers might make a decision for punish and pardon according to different mechanisms. 
The other is that, similar to predicting decisions, there are some commonalities in important variables across the category of toxic behavior for predicting overwhelming majority pardon and punish.
For example, in predicting overwhelming majority pardon, the most important variable in the report only model is the number of reports by allies across the category.  
Similarly, in predicting overwhelming majority punish, the most important variable in the chat only model is the number of messages sent by the offender across the categories.
Of course, there are some specifics for each category.
For predicting overwhelming majority punish, in the report only model, the number of reports by enemies is more important than the number by allies in intentional feeding, but in verbal abuse, allies' reports are more important than enemies'.
For future work, we intend to combine this result with qualitative user interviews and plan to reveal details of the mechanism of reviewers' decisions.

\begin{figure}[t!]
	\centering
	\includegraphics[width=\columnwidth]{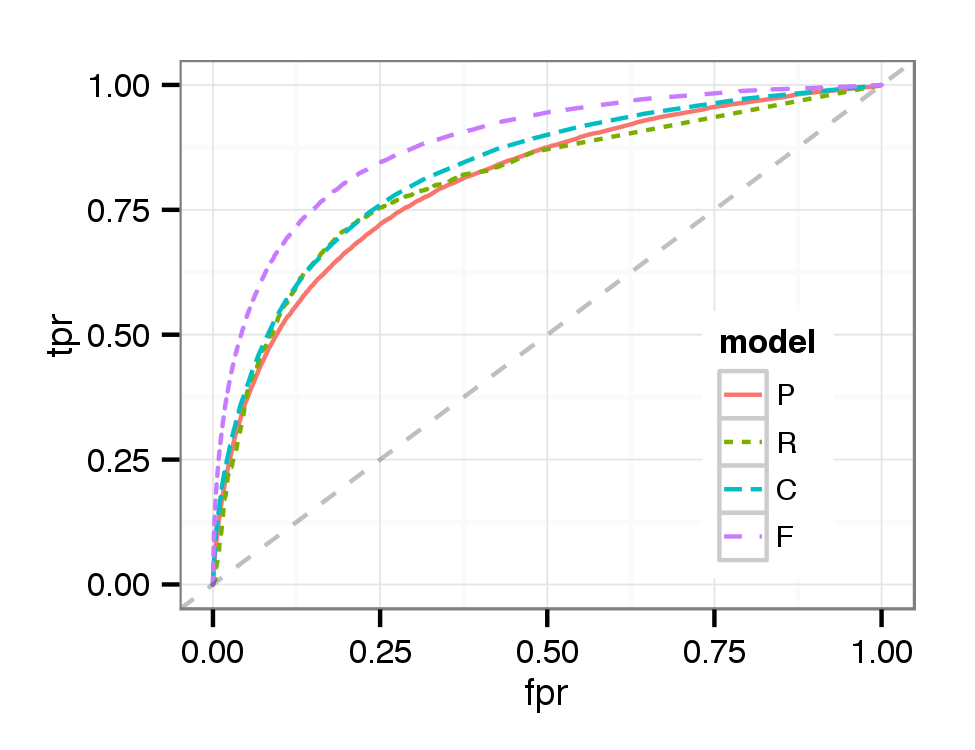}
	\caption{[Best viewed in color.] ROC curve for predicting overwhelming pardon decisions with models using in-game performance (P), user report (R), chats (C), and using all available features (F)}
	\label{fig:offender-allies-enemies-om-pardon}
\end{figure}

\begin{figure}[t!]
	\centering
	\includegraphics[width=\columnwidth]{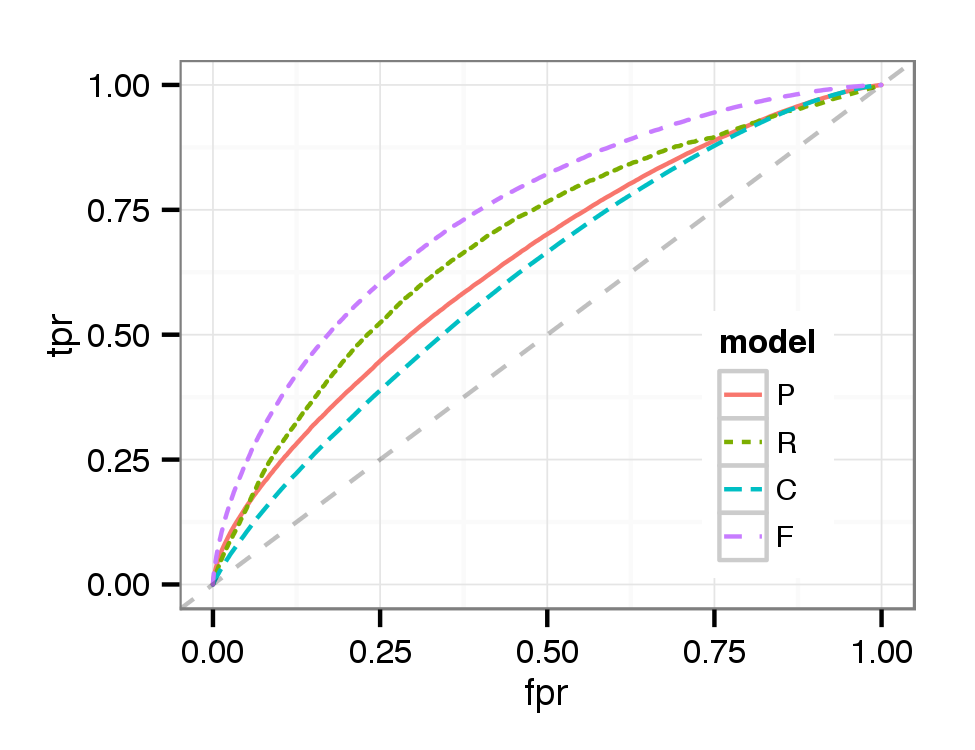}
	\caption{[Best viewed in color.] ROC curve for predicting overwhelming punish decisions with models using in-game performance (P), user report (R), chats (C), and using all available features (F)}
	\label{fig:offender-allies-enemies-om-punish}
\end{figure}

Figures~\ref{fig:offender-allies-enemies-om-pardon} and \ref{fig:offender-allies-enemies-om-punish} show ROC curves for predicting overwhelming pardon and overwhelming punish, respectively.  
Their AUC are 0.8049, 0.8055, 0.8269, and 0.8811 for overwhelming pardon decisions and 0.6509, 0.6886, 0.6190, and 0.7461 for overwhelming punish decisions.
There are some interesting differences between the curves for the agreement levels.  
First, detecting overwhelming pardon is easier to find than overwhelming majority punish and shows quite good performance.
It is mainly because overwhelming majority punish is very close to strong majority punish, as we mentioned in Figure~\ref{fig:majority-roc}.  

This proves the feasibility of automatically assigning tasks to crowds and machines according to their difficulties.  
Quinn et al. demonstrate that crowd-assisted machine learning can achieve high overall accuracy when assigning easy tasks to machine and fuzzy tasks to human~\cite{quinn2010crowdflow}.
Although they divide cases into two classes by human experts, our result demonstrates that we can do it automatically.

In context of LoL, properly dealing with overwhelming pardon case is more important than overwhelming punish.
Wrongly punished players would leave LoL, while wrongly pardoned players sometimes would be back to the Tribunal.
If they do not come to the Tribunal again, it means that they are reformed and fine for the overall LoL ecosystem. 

In addition, we achieve high accuracy to predict decisions of punish or pardon on clear-cut cases, overwhelming majority punish and pardon cases, as in Figure~\ref{fig:majority-roc}.
Thus, it is feasible that our classifier can automatically extract clear-cut cases and make accurate decisions on them.
It is great way to assist crowdsourcing platform by machine learning.  

Second, the order of models by performance is different in two cases.
In detecting overwhelming majority pardon, we observe that a chat model shows the best performance, while a user report model is quite comparable for the most part.  
By contrast, in detecting overwhelming majority punish, a user report model shows quite good performance.
This is an interesting finding.  
Reviewers need to understand context from chats to prove not guilty, but they also see why and how many times a toxic player is charged.  
This is consistent with our initial analysis, revealing the number of user reports is highly correlated with the likelihood of being punished~\cite{chisubmission}.

\subsection{Classifier portability}

\begin{figure}[t!]
	\centering
	\includegraphics[scale=0.24]{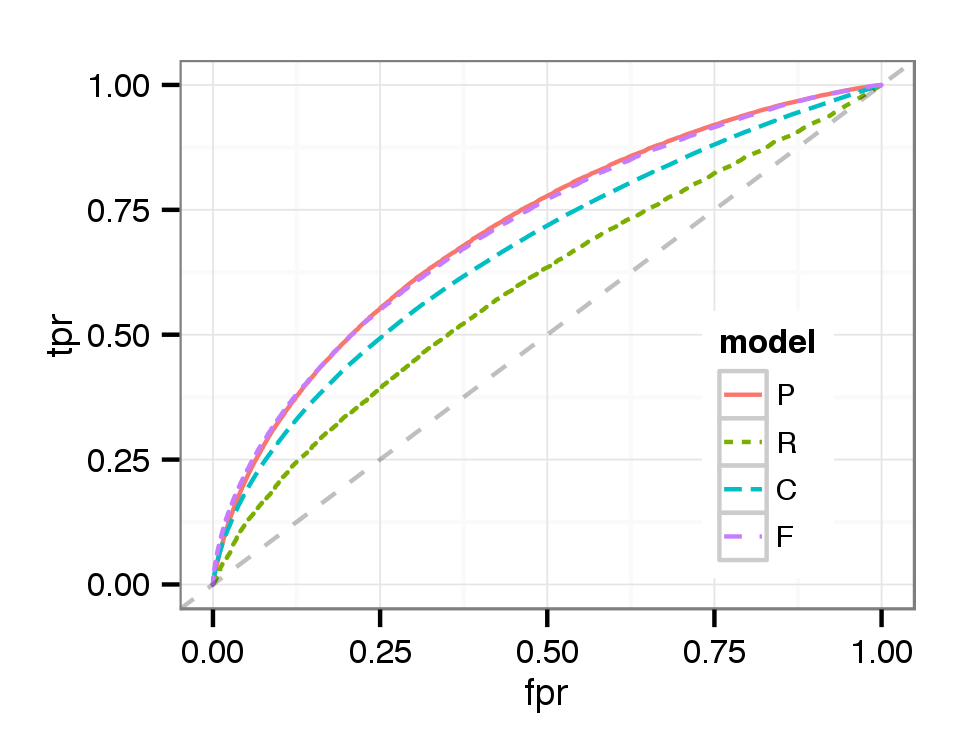}
	\caption{[Best viewed in color.] ROC curve for EUW decisions with classifier trained on NA.}
	\label{fig:portability_decisions}
\end{figure}

\begin{figure}[t!]
	\centering
	\includegraphics[scale=0.24]{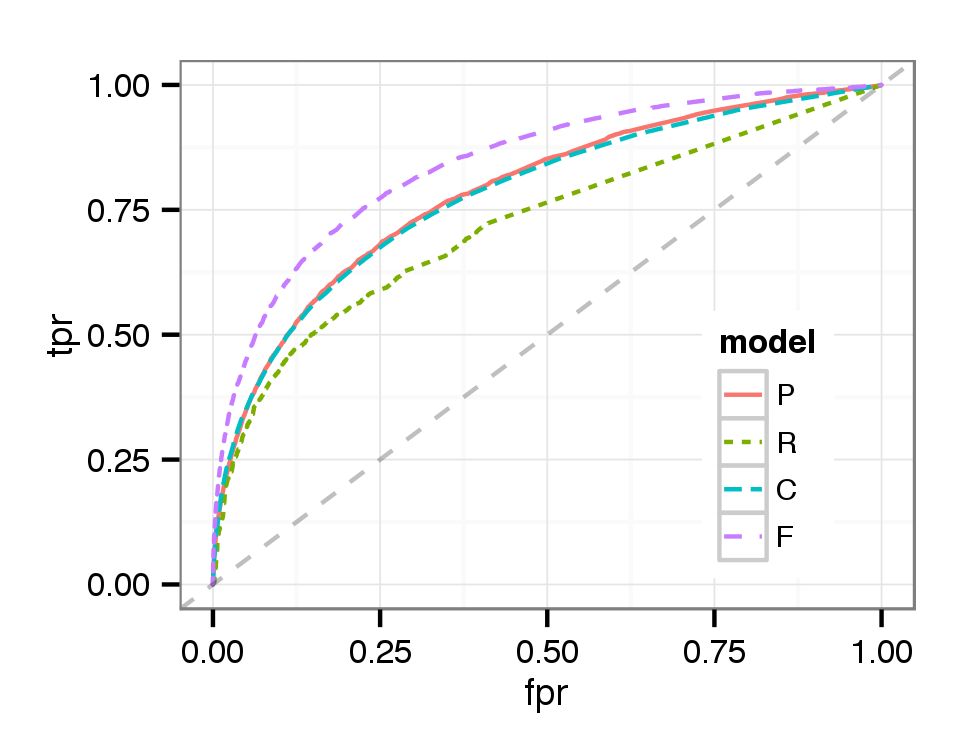}
	\caption{[Best viewed in color.] ROC curve for EUW Overwhelming Majority Pardons with classifier trained on NA.}
	\label{fig:portability_om_pardon}
\end{figure}

Finally, we explore whether or not our classifier is \emph{portable}.
Based on previous analysis~\cite{chisubmission}, we saw that there were statistically significant differences in Tribunal cases across the various regions that LoL operates.
One underlying reason behind this is likely due to cultural differences realizing themselves both in the tendencies of toxic players as well as the reviewers.
Because of these differences, we expect models trained on the North American dataset to not perform as well on the other regions.
However, we \emph{do} expect the models to remain superior to a coin flip in terms of discriminatory power.
In other words, while we believe the models we specify are meaningful regardless of the region, the thresholds learned are probably sub-optimal.

Before presenting results, we stress an additional caveat related to linguistic features.
The ANEW dataset is based on \emph{English} words and was built from \emph{American} respondents.
This makes the linguistic features (highly relevant in the NA dataset) useless when applied to the KR dataset since English language words are almost non-existent.
The EUW dataset includes players with a variety of native tongues, and anecdotally French, German, and Spanish are all spoken in-game.
However, there is no language requirement to become a reviewer; you only have to have an account within the region you are reviewing.
Also, English is a common tongue for gamers world wide.
In fact, we see that less than 1\% of EUW cases have an undefined $v_{text}$.

Figures~\ref{fig:portability_decisions} and \ref{fig:portability_om_pardon} show ROC curves of predicting EUW decisions and detecting EUW overwhelming majority pardon cases by using classifier trained on NA.  The performance of predicting EUW decision does not reach that of NA decision, but detecting EUW overwhelming majority pardon is as good as NA. 
As with our hypothesis, this shows the potential of classifier portability, but at the same time, the existence of regional differences~\cite{chisubmission}.

\section{Discussion}
\label{sec:discussion}

\subsection{Estimating the Real-world Impacts}

We present the potential gain of time, cost, and accuracy if our classifier assists the Tribunal.  One challenge is estimating the actual cost, time, and accuracy of reviewing cases in the Tribunal because Riot does not release detail statistics thereof, except a few infographics.
We collect and complement partially available information to estimate required sources for the Tribunal.  

First, we estimate the actual cost for crowdsourcing decisions.  Initially, Riot gave 5 Influence Points (IP) as rewards to each vote only when a majority vote is reached, but have since removed this payment system.  
In LoL, IP is used for buying champions or skins.
To measure how big 5 IP is, we need to convert it to real money.
Some champions whose price are 450 IP can also be bought for 260 Riot Points (RP), that can be purchased by real money.  
Players pay \$10 for 1380 RP.
Through a simple calculation, we reach \$0.02 for each correct vote.
This is comparable fare with other crowdsourcing platforms~\cite{quinn2010crowdflow}.

Second, we estimate the time required for each case.
According to the talk by Jeffrey ``Lyte'' Lin at Game Developers Conference (GDC) 2013\footnote{http://tinyurl.com/stfunub8}, reviewers have cast 105 million votes and reformed 280,000 toxic players.
Other announcements by Steve ``Pendragon'' Mescon\footnote{http://tinyurl.com/stfunub9} reveal 50\% of players warned by Tribunal are reformed.
We thus assume that 105 million votes make verdicts for 560,000 toxic players and half of them are reformed.
We conservatively assume this is the lower bound of players who came to the Tribunal.
This means that 187.5 votes are required for majority votes on a single case.
From the same source, Riot reveals more than 47 million votes were cast in the first year of the Tribunal, implying that 1.49 votes per second are cast.
From both, we can infer 125.85 seconds are required to reach a verdict for one case in the Tribunal.
We reasonably assume that this is the acceptable speed where Riot's in-house experts manually review some intriguing cases.

Finally, we estimate the accuracy of the Tribunal.  Lin said, ``approximately 80\% agreement between the [Tribunal] community and Riot's in-house team'', in the same GDC talk.
He added that the in-house team is less lenient than the Tribunal.
Surprisingly, the overall accuracy of the Tribunal is comparable with our classifier with respect to Riot's in-house decisions.  
That is, in contrast to CrowdFlow~\cite{quinn2010crowdflow}, our supervised learner has the potential to replace the crowdsourcing part of the Tribunal with no real sacrifice in accuracy. 

We now estimate the gain of Riot from the view of cost and victim players.

\textbf{Cost savings} ~~~ We already compute that the cost of each correct vote is \$0.02.  Conservatively, we estimate 50\% of all votes fall into majority for each case.
Since the Tribunal got 47 million votes the first year, its cost is 47M (votes) x 50 (\%) x 0.02 (\$) = 470,000 (\$). 
As of March 2013, the number of votes reached 105 millions.
Its potential cost surpasses 1 million dollars.
With the success of new regions and a growing userbase, this cost will become huge. 

\textbf{Protecting victims} ~~~ As of October 2012, Riot announced that 12 million players play LoL everyday\footnote{http://tinyurl.com/stfunub10} and they play more than 1 billion hours every month.
Thus, we estimate that a player enjoys LoL 83 minutes everyday which equates to 2.21 matches where one match usually spans 30 to 45 minutes.  
In the first year, the Tribunal detected 47M / 187.5 = 250,667 toxic players.
On average, 686.75 toxic players are warned by the Tribunal everyday.
From this number we can compute number of potential victims who are protected by the Tribunal everyday. 
The number of matches toxic players participate everyday is 1,517.74, and 13,659.61 innocent players are exposed to toxic players.
If our classifier and the Tribunal works together in a 50-50 manner, we can protect 13,659 more players everyday and more than 400 thousand per month.   

\subsection{Limitations and Consequences}

Although we have shown our approach is relatively robust, even across different regions of the world, there are some limitations.
First, although outside the scope of this paper, LoL features an ever changing ``meta-game'' which dictates commonly used strategies and tactics.
Although the features we use in this paper are not directly tied to the meta, for example which particular items or champions are selected, they are indirectly related.
E.g., the amount of damage dealt and received or gold earned might be influenced by the current meta.
Although the datasets in this paper span multiple metas and we still see good results, accuracy might be improved by training classifiers on data only from the meta of cases under examination.
We also note that changes in meta only affect the performance related features/models.

Toxic players could, in theory, adapt to avoid detection.
For example, toxic players might say ``Wow, you are a great player!'' sarcastically instead of calling someone a noob.
Or, perhaps an intentional feeder would go out of his way to damage the enemy team, but not enough to actually kill them.
This raises some interesting points.
First, toxic behavior only has an impact if it actually affects people in a negative way.
It would take a major shift in the community's understanding of language to infer that a seemingly positive statement was meant in a negative way.

Second, in the case of toxic players adapting their play style, we argue that is a hidden benefit of our approach.
Detecting toxic behavior has significant value, but preventing it wholesale or reducing its impact is a much better solution.
Although a feeder could attempt to hide his intentions by damaging the enemy team, he is consequently reducing the negative impact of his feeding by still providing some utility to his team.

\subsection{Future Research Directions}

One interesting research direction is real-time detection of toxic playing and adaptive interface design. 
Currently, the Tribunal deals with toxic players \emph{after} victims are exposed to online violence.
We quantitatively show that many people are exposed to toxic behavior even though a relatively well designed crowdsourcing framework to deal with toxic players is operating.
Our suggestion is to protect innocent players before they are exposed to toxic playing through adaptive an user interface.
Simply, we can warn toxic players when they exhibit toxicity.
Since toxic players sometimes do not recognize what they did as toxic playing~\cite{Lin05}, some visual cues representing toxicity will be helpful to reform them.  

As another psychological experiment, we are interested in whether they can be treated by giving penalties within a game, such as degrading champion status in real time, when toxic playing is detected.  
Since the competitive nature of LoL is the origin of some toxic playing, in-game penalties might have an impact.
The other reason that this might work is due to the immediacy of punishment.  
It is well known that instant punishment (reinforcement) is more effective than delayed one~\cite{van1983punishment}.
Degrading champion status or marking them visibly~\cite{Blackburn12} as toxic players could serve as an intermediate, direct punishment, while warning by the Tribunal is a delayed punishment. 
How toxic players react to this intermediate punishment will be an interesting avenue of future research.

\section{Summary}\label{sec:summary}

In this paper we explored the use of crowdsourced decisions on toxic behavior made by millions of experts.
Using the same sparse information available to the reviewers, we trained classifiers to detect the presence, and severity of toxicity.
We built several models oriented around in-game performance, reports by victims of toxic behavior, and linguistic features of chat messages.
We found that training with high agreement decisions resulted in more accuracy on low agreement decisions and that our classifier was adept in detecting clear cut innocence.
Finally, we showed that our classifier is relatively robust across cultural regions; our classifier built from a North American dataset performed adequately on a European dataset.

Ultimately, our work can be used as a foundation for the further study of toxic behavior.
The features we found to be important can serve as a spring board for more in depth learning.
Further, a preliminary cost analysis indicates that our classifier has the potential to, at minimum, alleviate the some of the burden placed on human reviewers.

%

\bibliographystyle{abbrv}
{
\small
\bibliography{arxiv-refs}  
}

%
%

\end{document}